\documentclass[aps,prd,twocolumn,showpacs,superscriptaddress,groupedaddress]{revtex4}

\usepackage{amssymb}
\usepackage{color,graphicx}
\usepackage{amsmath}
\usepackage{amsbsy}
\usepackage{amsthm}
\usepackage{bbm}
\usepackage{bm}
\usepackage{epsfig}
\usepackage{lscape}
\usepackage{float}
\usepackage{subfigure}
\usepackage{braket}
\usepackage{comment}
\usepackage{hyperref}
\usepackage[outdir=./]{epstopdf}

\usepackage[sort&compress]{natbib}

\newcommand{\com}[2]{\left[ {#1},{#2} \right]}

\newcommand{\tacom}[2]{\{ {#1},{#2} \}}
\newcommand{\bq}{\begin{equation}}
\newcommand{\eq}{\end{equation}}
\newcommand{\bqali}{\begin{equation}\begin{aligned}}
\newcommand{\eqali}{\end{aligned}\end{equation}}
\newcommand\erf{\operatorname{erf}}
\newcommand\besseli{\operatorname{I}}

\begin{document}
\author{Matteo Carlesso}
\email{matteo.carlesso@ts.infn.it}
\affiliation{Department of Physics, University of Trieste, Strada Costiera 11, 34151 Trieste, Italy}
\affiliation{Istituto Nazionale di Fisica Nucleare, Trieste Section, Via Valerio 2, 34127 Trieste, Italy}
\author{Angelo Bassi}
\email{bassi@ts.infn.it}
\affiliation{Department of Physics, University of Trieste, Strada Costiera 11, 34151 Trieste, 
Italy}
\affiliation{Istituto Nazionale di Fisica Nucleare, Trieste Section, Via Valerio 2, 34127 Trieste, Italy}
\author{Paolo Falferi}
\affiliation{Istituto di Fotonica e Nanotecnologie, CNR - Fondazione Bruno Kessler, I-38123 Povo, Trento, Italy}
\affiliation{INFN - Trento Institute for Fundamental Physics and Applications, I-38123 Povo, Trento, Italy}

\author{Andrea Vinante}
\email{anvinante@fbk.eu}
\affiliation{Istituto di Fotonica e Nanotecnologie, CNR - Fondazione Bruno Kessler, I-38123 Povo, Trento, Italy}

\title{Experimental bounds on collapse models from gravitational wave detectors}

\date{\today}

\begin{abstract}
Wave function collapse models postulate a fundamental breakdown of the quantum superposition principle at the macroscale. Therefore, experimental tests of collapse models are also fundamental tests of quantum mechanics.
Here, we compute the upper bounds on the collapse parameters, which can be inferred by the gravitational wave detectors LIGO, LISA Pathfinder and AURIGA. We consider the most widely used collapse model, the Continuous Spontaneous Localization (CSL) model. We show that these experiments exclude a huge portion of the CSL parameter space, the strongest bound being set by the recently launched space mission LISA Pathfinder. 
We also rule out a proposal for quantum gravity induced decoherence.
\end{abstract}

 \maketitle

\section{Introduction}

Wavefunction collapse models aim at solving the measurement problem of quantum mechanics, that is the contradiction between the linear and deterministic evolution of quantum systems and the nonlinear stochastic collapse of the wavefunction during a measurement process \cite{Ghirardi:1986aa,Bassi:2003ab,Bassi:2013aa}. The general assumption is that the quantum superposition principle breaks down at the macroscale due to a fundamental localization mechanism. In order to recover standard quantum mechanics at the microscale, the strength of the localization is assumed to be extremely weak at single particle level, while rapidly increasing with the number of constituents.

The most widely used collapse model is the so called Continuous Spontaneous Localization (CSL) model \cite{Ghirardi:1990aa,Ghirardi:1995aa}, which is based on two unknown constants, a characteristic length $r_C$, characterizing the spatial resolution of the stochastic collapse mechanism, and the collapse rate $\lambda$. The standard theoretical values commonly reported in the literature are respectively $r_C=10^{-7}$\,m, $\lambda=10^{-17}$\,s$^{-1}$ following Ghirardi, Rimini and Weber (GRW) \cite{Ghirardi:1986aa,Ghirardi:1990aa,Ghirardi:1995aa} and $r_C=10^{-7}$\,m, $\lambda=10^{-8\pm2}$\,s$^{-1}$ or $r_C=10^{-6}$\,m, $\lambda=10^{-6\pm2}$\,s$^{-1}$ following Adler \cite{Adler:2007aa}. These values are obtained when imposing in a somehow arbitrary way that macroscopic (in GRW case) or mesoscopic (Adler) quantum superpositions collapse in a reasonably short time. However, as the model is phenomenological, there is actually no fundamental way to predict the values of $r_C$ and $\lambda$. At present, GRW values can be regarded as a sort of lower limit, in the sense that weaker values of $\lambda$ would not guarantee a sufficiently rapid collapse of macroscopic human-scale quantum superpositions and this runs counter the original motivation of the model \cite{Toros:2016aa}.

Experimental tests of the CSL model can be done either with matter-wave interference experiments \cite{Hornberger:2012aa,Juffmann:2013aa,Arndt:2014aa,Marshall:2003aa,Wezel:2011aa,Romero-Isart:2011aa} or with noninterferometric methods \cite{Collett:2003aa,Adler:2005aa,Bahrami:2014aa, Belli:2016aa, Nimmrichter:2014aa, Diosi:2015ab, Goldwater:2015aa, Li:2016aa}. The strongest upper bounds so far have been set by the latter, in particular by X-ray spontaneous emission for $r_C<10^{-6}$\,m \cite{Curceanu:2015aa} and by force noise measurements on ultracold cantilevers for $r_C>10^{-6}$\,m \cite{Vinante:2016aa}. 

Here, we analyze the upper bounds that can be inferred by precision experiments based on macroscopic mechanical systems, focusing in particular on gravitational wave (GW) detectors. We will argue that GW detectors and related experiments, in particular the recently launched space mission LISA Pathfinder, set the strongest upper limits for $r_C>10^{-6}$\,m thereby excluding a huge portion of the CSL parameter space.
In section \ref{sec:theor} we will outline the theoretical model, in section \ref{sec:exp} we will derive the upper limit from three relevant experiments, respectively Advanced LIGO \cite{Abbott:2016ab}, LISA Pathfinder \cite{Armano:2016aa} and AURIGA \cite{Vinante:2006aa}. In section \ref{sec:disc} we will discuss the upper limits and compare them with the other existing bounds.

\section{Theoretical model}
\label{sec:theor}

Advanced LIGO, LISA Pathfinder and AURIGA represent the state of the art in their class, respectively ground-based interferometric detectors,  precursors of space-born detectors and resonant mass GW detectors. A GW detector monitors the deformation of space-time produced by gravitational waves. 
The strain noise spectrum $S_\text{\tiny hh}(\omega)$ quantifies the strength of such a deformation. 

\begin{figure}[ht!]
\centering
\includegraphics[width=0.6\linewidth]{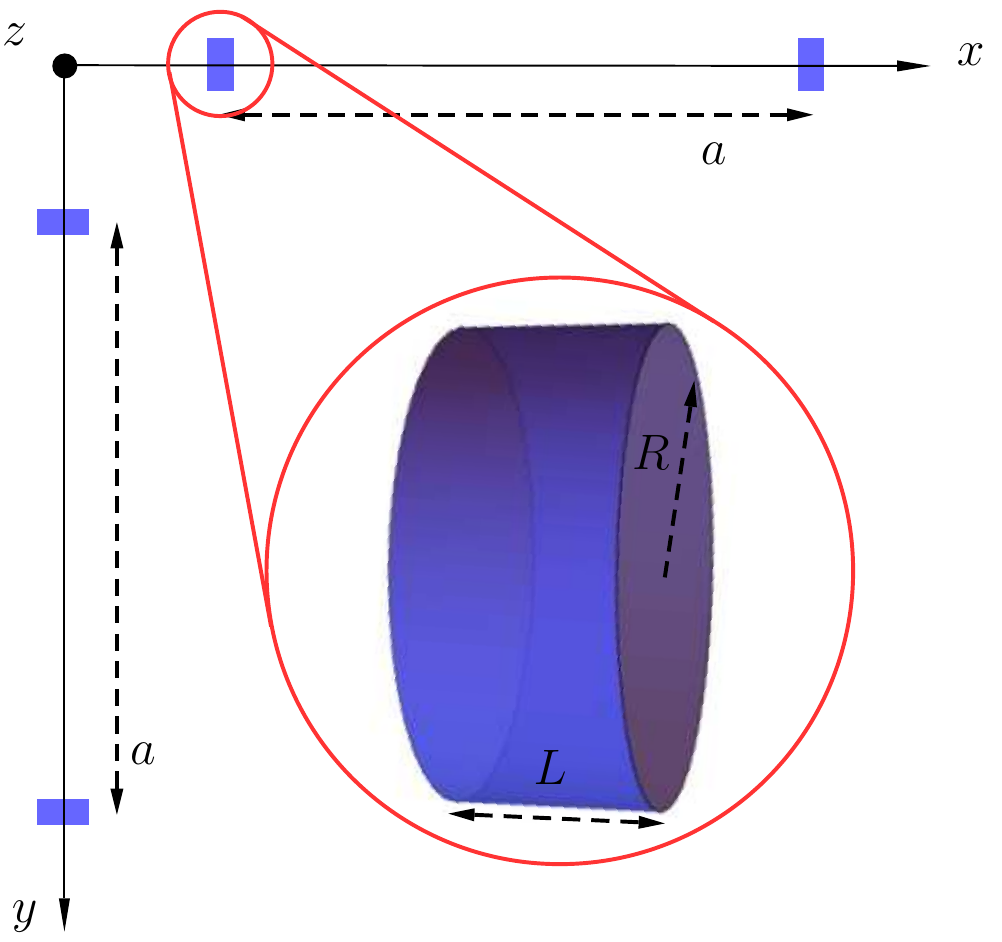}
\includegraphics[width=0.6\linewidth]{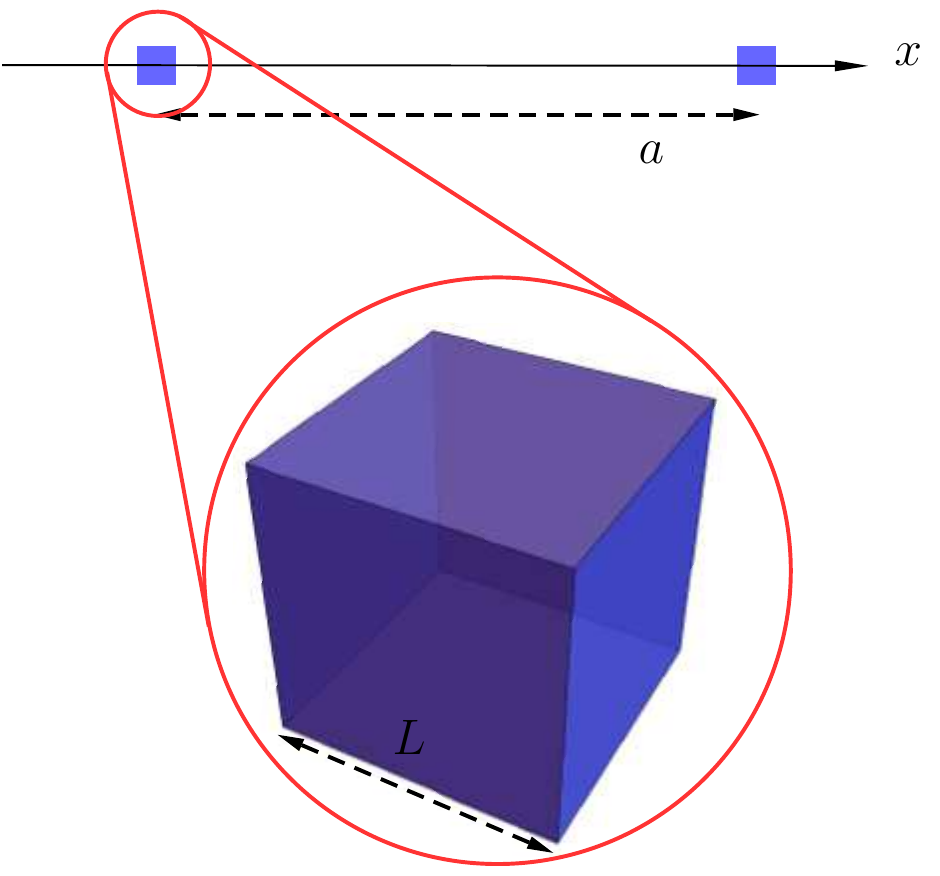}
\includegraphics[width=0.9\linewidth]{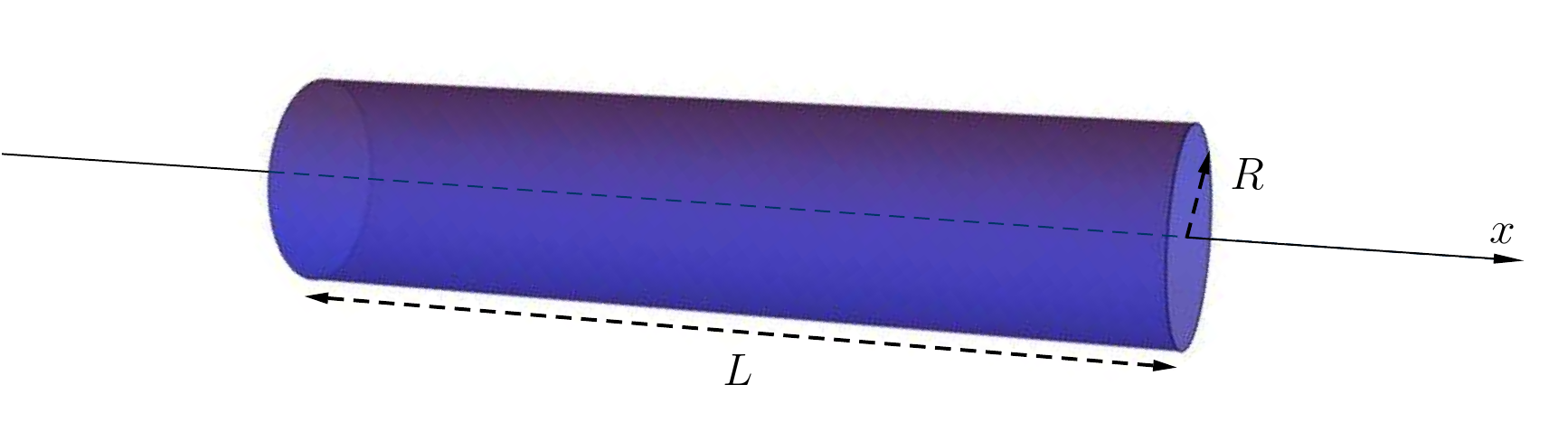}
\caption{(\textit{Color online}) Graphical representation of the three experiments here considered; the images are not in scale. LIGO on the top, LISA Pathfinder on the middle and AURIGA is on the bottom. In LIGO, four identical cylindrical masses (radius $R$, length $L$) are arranged as in Figure; $a$ is the distance between the center-of-mass of two masses on each arm of the interferometer. The arms are oriented along the $x$ and $y$ directions. LISA Pathfinder features two cubic (length $L$) masses, displaced along the $x$ direction with relative distance between their center-of-mass equal to $a$. AURIGA features a cylindrical single mass (radius $R$, length $L$), aligned with respect to the direction $x$ of measurement.}  \label{fig:geometria}
\end{figure}

Advanced LIGO and LISA Pathfinder monitor the optical distance between pairs of nominally free masses, while AURIGA is based on a single cylindrical bar mechanical oscillator (see Fig.~\ref{fig:geometria}). In the first case the CSL noise acts on the relative distance between the two masses; in the second case it causes a driving force on the bar oscillator. We will consider both cases.

The (mass proportional) CSL dynamics for the density matrix $\hat \rho(t)$ is
\cite{Bassi:2003ab}:
\bq\label{csl-eq}
\dfrac{d}{dt}\hat \rho(t)=-\dfrac{ \lambda}{2r_C^3\pi^{3/2}m_0^2}\int d{\bm z}\com{\hat M({\bm z})}{\com{\hat M({\bm z})}{\hat\rho(t)}},
\eq
where $m_0$ is a reference mass chosen equal to the mass of a nucleon, and $\hat M({\bm z})$ is defined as follows:
\bq\label{Mz-eq}
\hat M({\bm z})=m_0\sum_n e^{-\tfrac{({\bm z}-\hat {\bm q}_n)^2}{2r_C^2}},
\eq 
where the sum runs over the $N$ nucleons of the system; $\hat{\bm q}_n$ is the position operator of the $n$-th nucleon.

We divide the system in a subset of mass distributions, labeled by $\alpha$: for LISA Pathfinder two mass distributions ($\alpha=1$, 2), while for LIGO we have 4 mass distributions, but we will consider the two arms separately (so again $\alpha=1$, 2), for AURIGA we have a single cylindrical distribution ($\alpha=1$). Then, the position operator $\hat{\bm q}_n$ can be written as follows \cite{Belli:2016aa,Nimmrichter:2014aa}:
\bq
\hat{\bm q}_n={\bm q}^{(0)}_{n,\alpha}+\Delta \hat{\bm q}_{n,\alpha}+\hat {\bm q}_\alpha,
\eq
where ${\bm q}^{(0)}_{n,\alpha}$ is the classical equilibrium position of the $n$-th particle (belonging to the $\alpha$-th distribution), $\Delta \hat{\bm q}_{n,\alpha}$ measures the quantum displacement of the $n$-th particle with respect to its classical equilibrium position and 
$\hat {\bm q}_\alpha$ measures the fluctuations of the $\alpha$-th mass distribution. Under the assumption of rigid body, the latter fluctuations are the same for all the particles belonging to the $\alpha$-th distribution and therefore also for the $\alpha$-th center-of-mass, and $\Delta \hat{\bm q}_{n,\alpha}$ can be neglected.
When the spread of the center-of-mass wave-function is much smaller than $r_C$, Eq.~\eqref{Mz-eq} can be Taylor expanded up to the first order in $\hat {\bm q}_\alpha$:
\bq\label{Mz-lin}
\hat M({\bm z})\approx M_0({\bm z})+\sum_\alpha\int \dfrac{d{\bm x}}{r_C^2}\,\mu_\alpha({\bm x})e^{-\tfrac{({\bm z}-{\bm x})^2}{2r_C^2}}({\bm z}-{\bm x})\cdot\hat{\bm q}_\alpha,
\eq
where $M_0({\bm z})$ is a c-function, and $\mu_\alpha({\bm x})=m_0\sum_{n}\delta^{(3)}({\bm x}-{\bm q}_{n,\alpha}^{(0)})$ is the $\alpha$-th mass distribution. Here the sum runs on the nucleons belonging to the $\alpha$-th mass distribution. Eq.~\eqref{csl-eq} becomes
\bq\label{qumpl-eq}
\dfrac{d}{dt}\hat \rho(t)=-\dfrac{1}{2}\sum_{\alpha,\beta}\sum_{i,j=x,y,z}\eta^{\alpha,\beta}_{ij}\com{\hat q_{\alpha,i}}{\com{\hat q_{\beta,j}}{\hat \rho(t)}},
\eq
where $\hat q_{\alpha,i}$ is the $i$-th component of $\hat {\bm q}_\alpha$, and the diffusion CSL rate is given by
\bqali\label{etaj}
\eta^{\alpha,\beta}_{ij}&=\dfrac{\lambda}{r_C^7\pi^{3/2}m_0^2}\int d{\bm z}\int d{\bm x}\int d{\bm y}\, \mu_\alpha({\bm x})\mu_\beta({\bm y})\cdot\\
&\cdot e^{-\tfrac{({\bm z}-{\bm x})^2}{2r_C^2}}e^{-\tfrac{({\bm z}-{\bm y})^2}{2r_C^2}}({\bm z}-{\bm x})_i({\bm z}-{\bm y})_j.
\eqali
The dynamics in Eq.~\eqref{csl-eq} can be mimicked by a standard Schr\"odinger equation with an additional stochastic potential of the form
\bq
\hat V_\text{\tiny CSL}(t)=-\dfrac{\hbar\sqrt{\lambda}}{\pi^{3/4}r_C^{3/2}m_0}\int d{\bm z}\, \hat M({\bm z})w({\bm z},t),
\eq
where $w({\bm z},t)$ is a white noise with $\braket{w({\bm z},t)}=0$ and $\braket{w({\bm z},t)w({\bm y},s)}=\delta(t-s)\delta^{(3)}({\bm z}-{\bm y})$. 
Such a stochastic potential acts on the $\alpha$-th mass distribution as a stochastic force, which in the same limit of validity of the expansion in Eq.~\eqref{Mz-lin}, becomes
\bq\label{stochforce}
{\bm F}_\alpha(t)=\dfrac{\hbar\sqrt{\lambda}}{\pi^{3/4}m_0}\int  \dfrac{d{\bm z}d{\bm x}}{r_C^{7/2}}\,\mu_\alpha({\bm x})e^{-\tfrac{({\bm z}-{\bm x})^2}{2r_C^2}}({\bm z}-{\bm x})\,w({\bm z},t).
\eq
Notice that the noise $w({\bm z},t)$ is spatially uncorrelated: it acts randomly and independently on every nucleon of the system. However, the smearing function defined in the operator in Eq.~\eqref{Mz-eq} will introduce a spatial correlation, as we will see.

We now consider the $x$ direction of the motion of each mass distribution of the system, modelled as that of an harmonic oscillator of mass $m_\alpha$ and resonant frequency $\omega_\alpha$.
The corresponding quantum Langevin equations read:
\bqali\label{Langevin-eq}
\dfrac{d}{dt}\hat x_\alpha(t)&=\dfrac{\hat p_\alpha(t)}{m_\alpha}, \\
\dfrac{d}{dt}\hat p_\alpha(t)&=-m_\alpha \omega_\alpha^2 \hat x_\alpha(t) -\gamma_\alpha \hat p_\alpha(t) +F_\alpha(t),
\eqali
where $\hat p_\alpha(t)$ is the momentum of the $\alpha$-th distribution and $F_\alpha(t)$ is the stochastic force acting on it, both along the $x$ direction. We have added as usual a dissipative term $-\gamma_\alpha \hat p_\alpha(t)$, which can be expressed in terms of the quality factor of the system $Q_\alpha=\omega_\alpha/\gamma_\alpha$. A more general treatment should include additional noise terms to take into account the action of the environment and the measurement apparatus. However, since we are primarily interested in estimating the effect of the CSL noise, we neglect all other noise sources. Furthermore, the actual noise of the systems here considered is the sum of several noise sources (thermal, quantum, seismic, gravity gradient, etc.) and it is typically difficult to accurately distinguish and characterize each single contribution. This is typically the case for interferometric detectors. In order to set an upper limit on the CSL parameters we will take a conservative approach by assuming that all the experimentally measured noise is attributed to CSL. 
The physical quantity we are interested in is the 
force noise spectral density $S_{\text{\tiny FF}}(\omega)=\tfrac{1}{4\pi}\int_{-\infty}^{+\infty}\braket{\tacom{\tilde F(\omega)}{\tilde F(\Omega)}}$, expressed in N$^2$\,Hz$^{-1}$, where $\tilde F(\omega)$ is the Fourier transform of the $x$ component of stochastic force.

In the case of LISA Pathfinder and one arm of LIGO, there are two equal masses at an average distance $a$ and the monitored motion is the relative one, which is described by the following Langevin equations:
\bqali\label{Langevin-rel}
\dfrac{d}{dt}\hat x_\text{\tiny rel}(t)&=\dfrac{2\hat p_\text{\tiny rel}(t)}{m},\\
\dfrac{d}{dt}\hat p_\text{\tiny rel}(t)&=-\dfrac{m}{2} \omega_0^2 \hat x_\text{\tiny rel}(t) -\gamma \hat p_\text{\tiny rel}(t) +F_\text{\tiny rel}(t),
\eqali
where $F_\text{\tiny rel}(t)=\tfrac12 (F_1(t)-F_2(t))$. The corresponding force noise spectral density is given by
\bq\label{Sff-rel}
S_\text{\tiny FF}^\text{\tiny L}(\omega)=\dfrac{\hbar^2 \lambda r_C^3}{2\pi^{3/2}m_0^2}\int d{\bm k}\left|\tilde\mu({\bm k})\right|^2\left(1-e^{i a k_x}\right)k_x^2e^{-r_C^2{\bm k}^2},
\eq
where $\tilde\mu({\bm k})$ is the Fourier transform of $\mu({\bm x})$, and the correlation for the Fourier transformed white noise is $\braket{\tilde w({\bm z},\omega)\tilde w({\bm y},\Omega)}=2\pi\delta(\omega~+~\Omega)\delta^{(3)}({\bm z}-{\bm y})$. 
Here, there are two CSL contributions to the motion: the incoherent action on the single mass (first term in parenthesis) and the correlation between the two masses (second term), the latter being relevant when $a<r_C$. By substituting $\mu_{\alpha}({\bm r})$ with the $\alpha$-th mass distribution, a cylinder (radius $R$ and length $L$) for LIGO and a cube (length $L$) for LISA Pathfinder, we obtain from Eq.~\eqref{Sff-rel} the following expressions:
\begin{subequations}\label{SffL}
\bqali\label{SffLa}
S_\text{\tiny FF}^\text{\tiny LIGO}(\omega)&=\dfrac{8\hbar^2\lambda m^2}{L^2m_0^2}\left(\dfrac{r_C}{R}\right)^2\left(1-e^{-\tfrac{L^2}{4r_C^2}}+f_\text{corr}\right)\cdot\\
&\cdot\left[1-e^{-\tfrac{R^2}{2r_C^2}}\left(\besseli_0\left(\dfrac{R^2}{2r_C^2}\right)+\besseli_1\left(\dfrac{R^2}{2r_C^2}\right)\right)\right],
\eqali
\bqali
S_\text{\tiny FF}^\text{\tiny LISA}(\omega)&=\dfrac{16 \hbar^2\lambda m^2}{L^2m_0^2}\left(\dfrac{r_C}{L}\right)^4\left(1-e^{-\tfrac{L^2}{4r_C^2}}+f_\text{corr}\right)\cdot\\
&\cdot\left(1-e^{-\tfrac{L^2}{4r_C^2}}-\sqrt{\pi}\dfrac{L}{2 r_C}\erf\left(\dfrac{L}{2r_C}\right)\right)^2,
\eqali
\end{subequations}
where $\besseli_0$ and $\besseli_1$ denote the first two modified Bessel functions of the first kind, and $f_\text{corr}$ describes the correlations, which, because of the particular geometry of LIGO and LISA Pathfinder, have the same form:
\bq \label{dfgsdg}
f_\text{corr}=\dfrac12e^{-\tfrac{(a+L)^2}{4r_C^2}}\left(1+e^{\tfrac{aL}{r_C^2}}-2e^{\tfrac{L(2a+L)}{4r_C^2}}\right).
\eq
The effect of the correlations is to suppress the CSL effect in the relative motion of two equal masses when $r_C>a$.

In the case of LIGO, an extra factor 2 appears in Eq.~\eqref{SffLa} to take into account the two arms of the interferometer.

In the case of AURIGA, we have a single mass, and the monitored motion is the deformation of the resonant bar. The system can be modeled as two half-cylinders of mass $m/2$ and length $L/2$, connected by a spring and oscillating in counterphase with the same elongation of the bar extrema. The disposition of the two cylinders is the same  as that of the single arm of the LIGO experiment, with $a=L/2$ so that the two cylinders touch each other. 
The Langevin equations describing the relative motion of the two masses are described by Eq.~\eqref{Langevin-rel}, where $m/2$ replaces $m$ in the first of the two equations. Since the single arm of LIGO and our modeling of AURIGA have the same disposition,  Eq.~\eqref{SffLa} describes also the force noise spectral density for AURIGA, after replacing both $a$ and $L$ with $L/2$ in $f_\text{corr}$ (see Eq.~(\ref{dfgsdg})). The expression must also be divided by a factor 2 since there is only one arm. Eq.~\eqref{SffLa}  becomes:
\bqali\label{SffAURIGA2}
S_\text{\tiny FF}^\text{\tiny AURIGA}(\omega)&=\dfrac{4\hbar^2\lambda m^2}{L^2m_0^2}\left(\dfrac{r_C}{R}\right)^2\left(\frac32-\frac12e^{-\tfrac{L^2}{4r_C^2}}-e^{-\tfrac{L^2}{16r_C^2}}\right)\cdot\\
&\cdot\left[1-e^{-\tfrac{R^2}{2r_C^2}}\left(\besseli_0\left(\dfrac{R^2}{2r_C^2}\right)+\besseli_1\left(\dfrac{R^2}{2r_C^2}\right)\right)\right],
\eqali
where $m$ and $L$ are the mass and the length of the AURIGA cylinder. 

A final note: since the experimentally measured spectral densities refer only to positive frequencies, one has to multiply the expressions in Eq.~\eqref{SffL} and Eq.~\eqref{SffAURIGA2} by a factor 2 to take into account the conversion from the two-side to one-side spectra.

\section{Experimental upper bounds}
\label{sec:exp}

\subsection{Interferometric GW detectors: LIGO}

Interferometric GW detectors, such as LIGO \cite{Collaboration:2015ab} (as well as Virgo \cite{Acernese:2015aa}), are essentially Michelson interferometers in which the two arms are configured as a Fabry-Perot cavity. A passing gravitational wave induces a differential change of the arm lengths, resulting in a phase change of the output light. Each arm includes two suspended test masses acting as end mirrors, placed at several km to each other (4 km for LIGO, 3 km for Virgo) to maximize the response to the gravitational wave strain $h$. The suspensions are made of actively controlled multistage pendulum systems, with resonant frequency $\omega_0/2\pi$ below 1\,Hz, designed to heavily filter seismic noise. The last stage is designed for ultrahigh quality factor ($Q>10^8$) in order to suppress as much as possible the thermal noise. The actual frequency band sensitive to gravitational waves is roughly above $\sim 10$ Hz, implying that the test masses can be considered to a good approximation in the free-mass limit $\omega \gg  \omega_0$. 

Given the arm length $a$ (see Fig.~\ref{fig:geometria}), the differential change $\Delta a=|\Delta a_x -\Delta a_y|$ of the two arm lengths induced by an optimally oriented strain $h$ is predicted by General Relativity to be $\Delta a =h a$. It follows immediately that any displacement noise spectral density $S_{xx}^\text{\tiny LIGO}(\omega)$ of one of the two arms will cause an equivalent strain noise $S_\text{\tiny hh}^\text{\tiny LIGO}(\omega)=S^\text{\tiny LIGO}_{xx}(\omega)/a^2$. 
The former can be derived with the usual approach from Eqs.~\eqref{Langevin-rel} by solving them in the frequency domain \cite{Paternostro:2006aa}:
\bq\label{SxxLIGO}
S_{xx}^\text{\tiny LIGO}(\omega)=\dfrac{4}{m^2}\dfrac{S_\text{\tiny FF}^\text{\tiny LIGO}(\omega)}{(\omega_0^2-\omega^2)^2+(\frac{\omega\omega_0}{Q})^2},
\eq
where $S_\text{\tiny FF}^\text{\tiny LIGO}(\omega)$ is defined in Eq.~\eqref{SffL}.
In this way, in the free-mass limit $\omega\gg\omega_0$, we can derive the expression for the equivalent strain induced by the CSL noise. 

From Eq.~\eqref{SxxLIGO}, it follows that the CSL contribution to $S_\text{\tiny hh}(\omega)$ features a $1/\omega^4$ dependence, or a $1/\omega^2$ dependence when the square root spectrum $S_\text{\tiny h}(\omega)$ is considered.
The minimum force noise and therefore the strongest upper bound on the CSL parameters will be achieved at a well-defined frequency $\bar\omega/2\pi$. As the typically measured $S_\text{\tiny h}(\omega)$ is convex \cite {Collaboration:2015ab}, $\bar\omega/2\pi$ can be graphically inferred from the spectrum as the frequency at which $S_\text{\tiny h}(\omega)$, displayed in log-log scale, is tangent to a straight line with slope equal to $-2$.

For Advanced LIGO at the time of the first detection \cite{Abbott:2016ab, Abbott:2016aa}, $S_\text{\tiny h}(\bar\omega)$ is in the range of $10^{-23}$\,${\rm Hz}^{-\frac12}$. From the published spectrum we infer that the effective force noise reaches a minimum $S_\text{\tiny F}(\bar\omega)\approx 95 \,\rm fN \,\rm Hz^{-\frac12}$ at $\bar\omega/2\pi\sim 30-35$\,Hz. We have used the numerical values $m=40$\,kg for the test mass and $a=4$\,km for the arm length.
For the design sensitivity of Advanced LIGO, not yet reached, one can estimate from the design curves a minimum force $S_\text{\tiny F}(\bar\omega)\approx 25 \,\rm fN \, \rm Hz^{-\frac12}$ at $\bar\omega/2\pi \sim 15-20$\,Hz \cite{Abbott:2016ab, Abbott:2016aa, Collaboration:2015ab}.  Each test mass is a cylinder of fused silica (density = 2200\,kg/m$^3$) with radius $R=17$\,cm and length $L=20$\,cm.

By plugging the test mass parameters and the measured force noise in Eq.~\eqref{SffL}, we obtain the exclusion region for the CSL parameters shown in blue in Fig.~\ref{all}.
The achievable upper bound from the foreseen design sensitivity is shown with dashed blue line. 

\subsection{Space-based experiments: LISA Pathfinder}

The second system we consider is LISA Pathfinder. This space mission has been recently launched as a technology demonstrator of the proposed space-based gravitational wave detector LISA. LISA concept is similar to terrestrial interferometric detectors, but will exploit a much longer baseline $\sim 10^6$\,km and the more favourable conditions of operation in space. The detector will be sensitive to gravitational waves in the mHz range, thus providing different and complementary informations compared to ground-based detectors. LISA test masses will be in nearly ideal free-fall and essentially free from the vibrational, seismic and gravity gradient disturbances which unavoidably affect any terrestrial low-frequency experiment.

The main goal of LISA Pathfinder is to demonstrate the technology required by LISA, in particular to assess the accuracy of the achievable free-fall condition. The core of LISA Pathfinder consists in a pair of test masses (see Fig.~\ref{fig:geometria}) in free-fall, protected by a satellite which follows the mass trying to minimize the stray disturbance. The overall objective is to demonstrate the performance required for the test masses of LISA Pathfinder in terms of acceleration noise. Thus, the output of the experiment is directly expressed as a relative acceleration noise spectrum $S_\text{\tiny gg}(\omega)$, which is related to the relative force noise spectral density by the relation:
\bq\label{Sgg}
S^\text{\tiny LISA}_\text{\tiny gg}(\omega)=\dfrac{4}{m^2}S^\text{\tiny LISA}_\text{\tiny FF}(\omega).
\eq
The geometry of each test mass is straightforward: a cube of side $L=4.6$\, cm, made of an alloy of AuPt, with a mass $m=1.928$ kg, and the distance between the two masses $a=37.6$\,cm. Thanks to the space operation, it is possible to achieve a force sensitivity better than ground-based experiments. The current best experimental figure reaches a minimum acceleration noise of $S_\text{\tiny gg}(\omega)=2.7 \times 10^{-29}$\,m$^2$ s$^{-4}$ Hz \cite{Armano:2016aa}.

Using Eq.~\eqref{SffL} and Eq.~\eqref{Sgg}
and plugging in the numerical values of the parameters and the best force noise, as given above, we obtain the green exclusion area in Fig.~\ref{all}. 
 {The force noise in LISA Pathfinder is steadily improving with time likely because of progressive outgassing of the spacecraft, and is already significantly better than the published data \cite{Weber:private}. Assuming a reasonable improvement by factor of 2 we get the dashed green line in Fig.~\ref{all}.
Notice that this result would overcome the bound set by the ultracold cantilever experiment \cite{Vinante:2016aa} for the standard value taken for $r_C=10^{-7}\;\rm m$.}

\subsection{Resonant GW detectors: AURIGA}

The principle of resonant-mass GW detectors is to monitor the deformation of an elastic body, typically a massive ton-scale resonant bar or sphere, induced by a gravitational wave. The main drawback compared to interferometers, see above, is the smaller bandwidth and the shorter characteristic length $\sim $ 1\,m. However, as these detectors have been operated at cryogenic temperature and have achieved impressive displacement noise $\sqrt{S_{xx}}\sim $ 10$^{-20} \; \rm m \, Hz^{-\frac{1}{2}}$, it is worth considering their sensitivity to CSL effects. As best case we consider AURIGA \cite{Vinante:2006aa, Baggio:2005aa}, which is based on a aluminum (density = $2700$ kg/m$^3$) cylinder with length $L=3$\,m, radius $R=0.3$\,m and mass $m=2300$\,kg cooled to $T=4.2$\,K, schematically represented in Fig.~\ref{fig:geometria}. Other detectors of the same class feature similar parameters. 

The fundamental longitudinal mode of deformation at $\omega_0/2\pi\sim 900$ Hz is monitored by a sensitive SQUID-based readout \cite{Baggio:2005aa}. The system, as described above, is model as two masses $m/2$ connected by a spring and oscillating in counterphase. We expect this procedure to yield a crude but reasonable estimate of the CSL effect, within a factor of 2.
The equivalent force noise spectrum $S_\text{\tiny FF}(\omega)$ of the reduced system is related to the strain noise spectrum $S_\text{\tiny hh}(\omega)$ by the relation \cite{Vinante:2006aa, McHugh:2005aa}:
\bq
S^\text{\tiny AURIGA}_\text{\tiny FF}(\omega)  =\left(\dfrac{m \omega_0^2 L}{\pi^2}\right)^2S^\text{\tiny AURIGA}_\text{\tiny hh}(\omega).
\eq
For the AURIGA detector in the current scientific run, the minimum strain noise at resonance is $S_\text{\tiny h}(\bar \omega)=\sqrt{S_\text{\tiny hh}(\bar\omega)}=1.6 \times 10^{-21}$\,Hz$^{-\frac12}$ at $\bar\omega/2\pi =931$\,Hz (in the following we will use single index to represent square rooted spectral densities). An independent absolute calibration was performed, based on the fluctuation-dissipation theorem, demonstrating that the noise at resonance is dominated by thermal noise \cite{Vinante:2006aa}. The calibration accuracy was of the order of $\sim 10$\% in energy. Taking this into account, we estimate the minimum unknown force noise, which could be attributed to CSL, as $S_\text{\tiny F}=12 \, {\rm pN}\,{\rm Hz}^{ - \frac{1}{2}} $. Note that AURIGA (as well as NAUTILUS \cite{Astone:1997aa}) has been also operated in previous runs at lower temperatures $\sim 100$ mK \cite{Zendri:2000aa}. The minimum strain noise at resonance was actually lower, but an accurate thermal noise calibration in that case was not performed. This amounts to a value for the minimum unknown force noise comparable to that given above.

\begin{figure}[ht!]
\includegraphics[width=\linewidth]{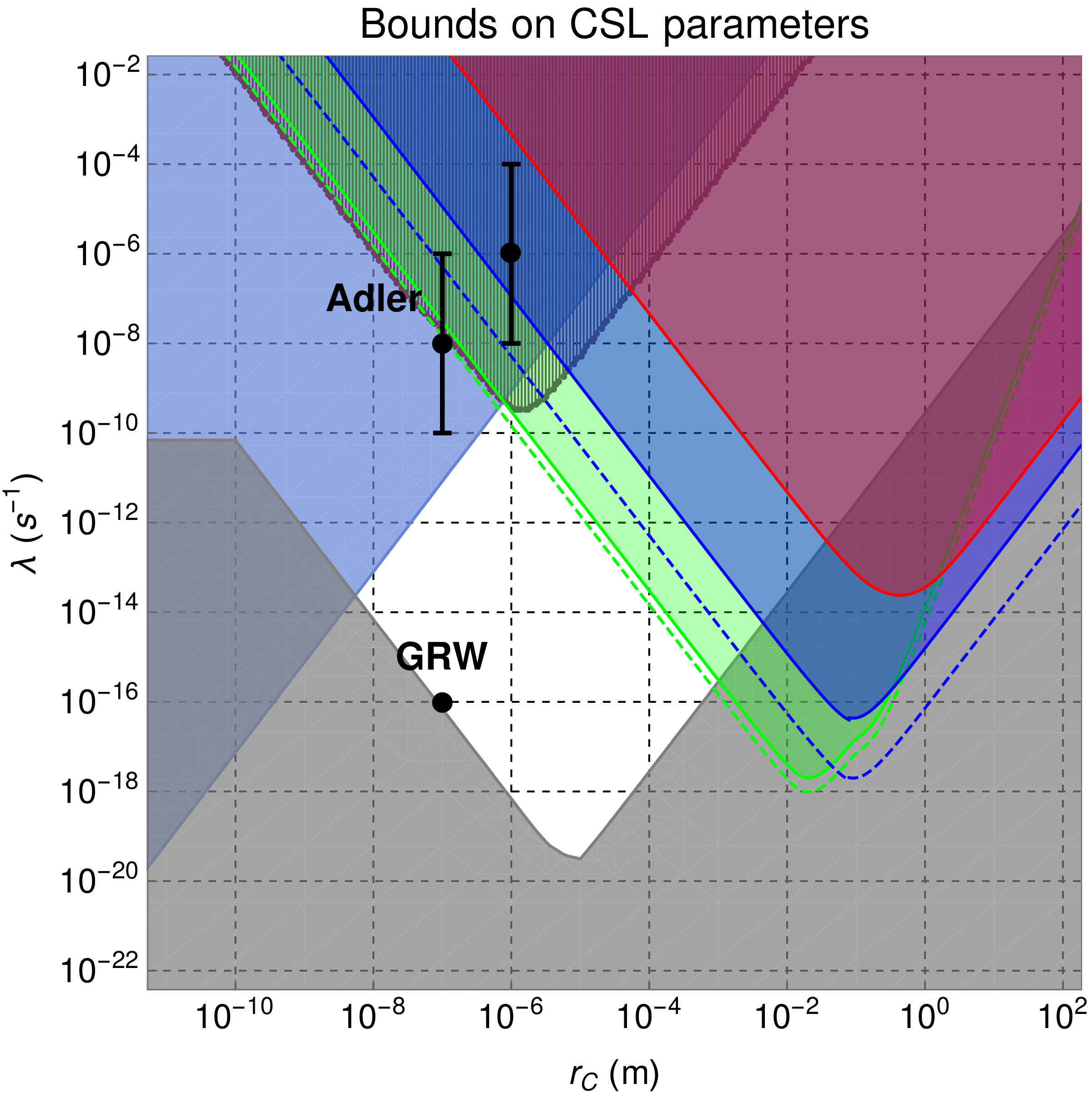}
\caption{(\textit{Color online}) Upper and lower bounds on the CSL collapse parameters $\lambda$ and $r_C$. Blue, green and red lines (and respective shaded regions): upper bounds (and exclusion regions) from LIGO, LISA Pathfinder and AURIGA.{ Blue and green dashed lines: upper bounds from foreseen improved sensitivity respectively of LIGO and LISA Pathfinder}. Purple line: upper bound from ultracold cantilever experiments \cite{Vinante:2016aa}. Light blue line: upper bound from X-ray experiments \cite{Curceanu:2015aa}. {Other weaker bounds \cite{Belli:2016aa,Bilardello:2016aa,Laloe:2014aa,Toros:2016aa,Toros:2016ab} are not reported. }Gray line: lower bound  based on theoretical arguments \cite{Toros:2016aa}. The GRW \cite{Ghirardi:1986aa,Ghirardi:1995aa} and Adler \cite{Adler:2007aa} values and ranges are indicated in black. }  \label{all}
\end{figure}

The comparison of the CSL prediction with the experimental data leads to the red line and exclusion area in Fig.~\ref{all}.

\section{Discussion}
\label{sec:disc}

The three exclusion regions computed here 
have a very similar shape, achieving a minimum for $r_C$ of the order of the test mass relevant length. For $r_C>1$\,m the bounds are roughly comparable, with the one set by LIGO slightly better. Theoretically, such values of $r_C$ are not much interesting, since already excluded by assuming the effective collapse of macroscopic objects \cite{Toros:2016aa} (gray region in Fig. 2). For smaller values of $r_C$, the best bound is set by LISA Pathfinder. 

We observe that the bounds are the best so far, for $r_C$ ranging from roughly 1 $\mu$m up to the macroscopic scale, thereby excluding a substantial region of the parameter space. While this $r_C$ interval is not the one usually considered relevant or theoretically favoured, we point out that, as long as $r_C$ and $\lambda$ are free parameters, unambiguous exclusion of a given region of parameter space can be done only through experiments.

At the standard characteristic length $r_C=10^{-7}$\,m, the bound from LISA Pathfinder $\lambda<3 \times 10^{-8}$\, s$^{-1}$ is still interesting, falling about a factor of 2 from the best limit obtained so far with mechanical techniques at ultralow temperature \cite{Vinante:2016aa}. The strongest bound so far for $r_C=10^{-7}$\,m is still provided by X-ray experiments \cite{Curceanu:2015aa}, although the latter requires stronger assumptions on the CSL noise spectrum.

We also note that the inferred bounds are conservative, at least for the LIGO and LISA Pathfinder cases, as we have assumed that all measured noise is attributed to CSL. Actually, the interferometer noise can be to a good extent characterized and attributed to well-defined sources. Subtraction of well-characterized noise may enable in principle a slight improvement of the bounds. For the AURIGA case, the noise at resonance is almost entirely due to thermal noise, and an absolute calibration based the fluctuation-dissipation theorem was performed. In this case a noise subtraction within the calibration uncertainty is entirely legitimate. For interferometers such as LIGO this task might be more difficult, as several noise sources combine together to yield the measured spectrum, depending on frequency. Some of them, such thermal noise, can be in principle fully characterized, but for others, like newtonian or seismic noise, the task is much more complicated. For the LISA Pathfinder case, there is evidence that thermal noise from the residual gas is dominating the residual force noise. Unfortunately, uncertainty in the pressure and composition of the gas make it hard to perform an independent calibration based on the fluctuation-dissipation theorem \cite{Armano:2016aa}.

We also mention that there is another class of macroscopic mechanical resonators, namely torsion balances, which have been the most sensitive force sensors since the time of Cavendish. We have not considered explicitly this class of experiments because typical sensor size ($10^{-2} - 1$ m) and frequency band (mHz) are very similar to the those of LISA Pathfinder. In fact, ground testing of LISA technology has been primarily done by means of torsion pendulum experiments. However, the actual performances achieved by LISA Pathfinder have arguably improved over ground-based tests by at least 2 orders of magnitude \cite{Armano:2016aa}.

As an immediate consequence of our analysis, we can also exclude a quantum gravity induced decoherence model proposed long time ago by Ellis \textit{et.~al.~}\cite{Ellis:1984aa,Ellis:1989aa}. Briefly, the model estimates the decoherence induced by the interaction with a background of wormholes. As long as the wavelength of the wormholes is much longer than the characteristic magnitude of the motion of the system, decoherence can be effectively described by the one-dimensional version of Eq.~\eqref{qumpl-eq}, with diffusion coefficient
\bq
\eta_\text{\tiny Ellis}=\dfrac{(c m_0)^4m^2}{(\hbar m_\text{\tiny Pl})^3},
\eq
where $m_\text{\tiny Pl}$ is the Planck mass and $c$ the speed of light. A recent analysis~\cite{Minar:2016ab} shows that this model is incompatible with the latest atom interferometry experiment of Kasevich's group, performing a spatial separation of $\sim0.5$\,m \cite{Kovachy:2015aa}. However in the latter case the negative result is not very strong, as the experimentally measured rate $\eta_\text{\tiny exp}$ is just $\sim 25$ times smaller than $\eta_\text{\tiny Ellis}$. In our case, data from LISA Pathfinder show that $\eta_\text{\tiny exp}$is $\sim 10^{12}$ times smaller than $\eta_\text{\tiny Ellis}$, thus setting a significantly stronger bound.

Finally, we discuss future prospects. For the present class of interferometers like Advanced LIGO or Advanced Virgo, a significant improvement is expected in the next 2-3 years, with these detectors likely approaching the design sensitivity. 
Blue dashed line in Fig.~\ref{all} shows the bound from the design sensitivity of LIGO. 
The bound would not improve over the LISA Pathfinder one at short $r_C$, but would further extend the exclusion region at $r_C>1$ m. The third generation of interferometers, currently under study, will employ cryogenic suspensions for a further reduction of low frequency noise by 1-2 orders of magnitude. This may enable an improvement of the CSL bounds over LISA Pathfinder. Inversely, we note that if a CSL noise will eventually appear in the range of parameters predicted by Adler, this would limit the low-frequency sensitivity of future generation of interferometers. While this scenario might seem unlikely, it seems it was never clearly pointed out.

On the other hand, strong improvements are expected by space missions in the near and far future. LISA Pathfinder is still under operation and the noise is slowly but steadily improving with time, likely due to slowly decreasing gas pressure \cite{Armano:2016aa}. Within the next months the noise is expected to further improve and strengthen the bound, as anticipated in Fig.~\ref{all}. On the other hand, LISA Pathfinder has essentially reached the requirements for the future LISA mission in terms of residual acceleration noise. While no substantial improvement is required by LISA, it is reasonable to expect further progress in the next decade before the launch. Other missions under study, such as MAQRO, will try to exploit the beneficial aspects of space environment in interferometric or force-sensing experiments with nanoparticles with size around $10^{-7}$ m \cite{Kaltenbaek:2012aa}. This may open the way to a full test of CSL and other collapse models in a more relevant range of parameters.
\\

\noindent \textit{Note.} Preliminary results presented at 115th Statistical Mechanics Conference, Rutgers University 8-10 May 2016, and at the Quantum Control of Levitated Optomechanics Conference, Pontremoli, 18-20 May 2016. During the completion of this paper, we became aware of a related work by B. Helou {\it et al.} \cite{Helou:2016aa}, deriving upper limits on collapse models from LISA Pathfinder data. 

\section*{ACKNOWLEDGEMENTS}

MC and AB acknowledge support from the University of Trieste and INFN. AV thanks S. Vitale, K. Danzmann and W.J. Weber for discussions on LISA Pathfinder data.

\end{document}